\documentstyle[11pt,aasms4]{article}
\def\alwaysmath#1{\ifmmode{#1}\else{$#1$}\fi}

\def\arcsec{\hbox{$^{\prime\prime}$}}

\def\ers{\alwaysmath{{\rm \, erg\,sec^{-1}}}}

\righthead{Ferraro, et al.}
\lefthead{}

\slugcomment{In press in the Astrophysical Journal Letters}

\begin{document}

\title{Very Large Telescope Observations of the peculiar globular cluster
NGC\,6712. 
Discovery of a UV, H$\alpha$ excess star in the core\altaffilmark{1}}
 
\author{Francesco R. Ferraro\altaffilmark{2},
	Barbara Paltrinieri\altaffilmark{3,4},
	Francesco Paresce\altaffilmark{5},
	Guido De Marchi\altaffilmark{5,6,7}}
 
\altaffiltext{1}{Based on observations collected at ESO-VLT, Cerro
Paranal, Chile}
\altaffiltext{2}{ Osservatorio Astronomico di Bologna, via Ranzani 1,
I--40126 Bologna, Italy; ferraro@bo.astro.it.}
\altaffiltext{3}{Istituto di Astronomia --- Universit\'a La sapienza,
Via G.M. Lancisi 29, I--00161 Roma, Italy;
barbara@coma.mporzio.astro.it} 
\altaffiltext{4}{Osservatorio Astronomico di Roma, Via Frascati, 33, 
I--00040 Monteporzio Catone, Italy}
\altaffiltext{5}{European Southern Observatory, Karl Schwarzschild
Strasse 2, D--85748 Garching bei M\"unchen, Germany; fparesce@eso.org}
\altaffiltext{6}{Space Telescope Science Institute, 3700 San Martin
Drive, Baltimore, MD 21218, USA; demarchi@stsci.edu}
\altaffiltext{7}{On assignment from the Astrophysics Division, Space
Science Department of ESA}
\begin{abstract}

We present results from  multi-band observations in the central region
of the cluster NGC\,6712 with the ESO {\it Very Large Telescope}. Using
high resolution  images we have identified three UV-excess stars. In
particular two of them are within the cluster core, a few arcsec
apart:  the first object is star {\it "S"} 
which previous studies identified as  
the best candidate to the optical counterpart to the luminous X-ray
source detected in this cluster.
 The other UV object  shows clearcut
$H\alpha$ emission  and, for this reason, is an additional promising
interacting binary candidate (a quiescent LMXB or a CV). The presence
of two unrelated interacting binary systems a few arcsec apart in the
core of this low-density cluster is somewhat surprising and supports
the hypothesis that the (internal) dynamical history of the cluster
and/or the (external) interaction with the Galaxy might play a
fundamental role in the formation of these peculiar objects.
\end{abstract}


\section{Introduction} \label{sec:intro}
  
Two main classes of X-ray sources have been found to exist in Galactic
globular clusters (GGCs):  {\it (i)} high luminosity X-ray sources
(with $ L_{X} > 10^{34.5} \ers$), the so-called Low Mass X-ray
Binaries, LMXB; {\it (ii)} low luminosity X-ray sources (hereafter
LLGCXs) with $ L_{X} < 10^{34.5} \ers$).  Although the true nature of
these objects is still a matter of debate,  both these categories of
objects are thought to be associated with interacting binary systems.
In particular, LMXBs, because of their X-ray bursts, might  be binary
systems with an accrete neutron star, while LLGCXs (or at least the
faintest LLGCXs with $ L_{X} < 10^{32} \ers$) are supposed to be
binary systems in which a white dwarf (instead of a neutron star) is
accreting material from a late type dwarf (main sequence or sub-giant
branch star; see Verbunt et al. 1994), and, for this reason, they
might be possibly connected to cataclysmic variables (CVs).

In  dense environments such as the cores of GGCs one can expect
interacting binaries to result from the evolution of various kinds of
binary systems, with a variety of origins and nature. For example,
binaries in dense clusters could have been created by dynamical
processes (Hut \& Verbunt 1983; Bailyn 1995), while  in low-density
clusters they might result from the evolution of primordial systems
(Verbunt \& Meylan 1988).

In this scenario, the case of NGC\,6712 is quite interesting. Although
it is a relatively loose, intermediate density GGC ($c=0.9$, $Log
\rho_0 \sim 3$;, Djorgovski \& Meylan 1993), it is one of the 12 (out of
150) GGCs harbouring in its core a LMXB burster ($1RXS
J185304.8-084217$; see Voges et al. 1999).  Indeed, NGC\,6712 is the
lowest density cluster in the Galaxy containing a LMXB: its central
density is significantly lower than the mean density of  GGCs
containing LMXBs ($<Log \rho_0^{LMXB}> =4.9\pm0.3$, Bellazzini et al.
1995).
  
The optical counterpart to the X-ray burster in NGC\,6712 has long been
searched (see Cudworth 1988, Bailyn et al. 1988, Nieto et al. 1990).
Finally, it has been identified by ground based observations
as a faint UV-excess star (star $S$) by
Bailyn et al. (1991) and  Auriere \& Koch-Miramond (1992) and afterwards
 confirmed from HST observations by Anderson et al. (1993).

Moreover, NGC\,6712 shows another peculiar characteristic:  its orbit,
as computed by Dauphole et al. (1996), suggests that this cluster has
experienced a severe interaction with the Galaxy during its numerous
passages through the disk and bulge. Preliminary VLT observations (De
Marchi et al., 1999), now confirmed by Andreuzzi et al. (2000), fully
support this picture, since the main sequence (MS) luminosity function
(LF) shows that the mass fuction of the cluster has been severely
depleted of lower  mass stars, probably  stripped by  the tidal force
of the Galaxy.
 
As part of a long term project specifically devoted to the study of
the global stellar population in a sample of GGCs, we imaged the core
of NGC\,6712, exploiting the exceptional performances of the ESO {\it
Very Large Telescope}. The complete data set, together with other
specific aspects connected to the unevolved  (MS) and evolved (red
giant branch, horizontal branch, etc.) stars, are  discussed separately
in a series of companion papers (Andreuzzi et al. 2000, Paltrinieri et
al. 2000, in preparation).  In this letter, we briefly discuss the
properties of a few UV-excess stars, in particular 
 we report on the discovery of an
object  with   strong UV and H$\alpha$ excess, located in the very
central region of NGC\,6712, a few arcsec away from the optical
counterpart to the LMXB source.

\begin{figure*}[htb]
\vskip4.5truein
\includegraphics{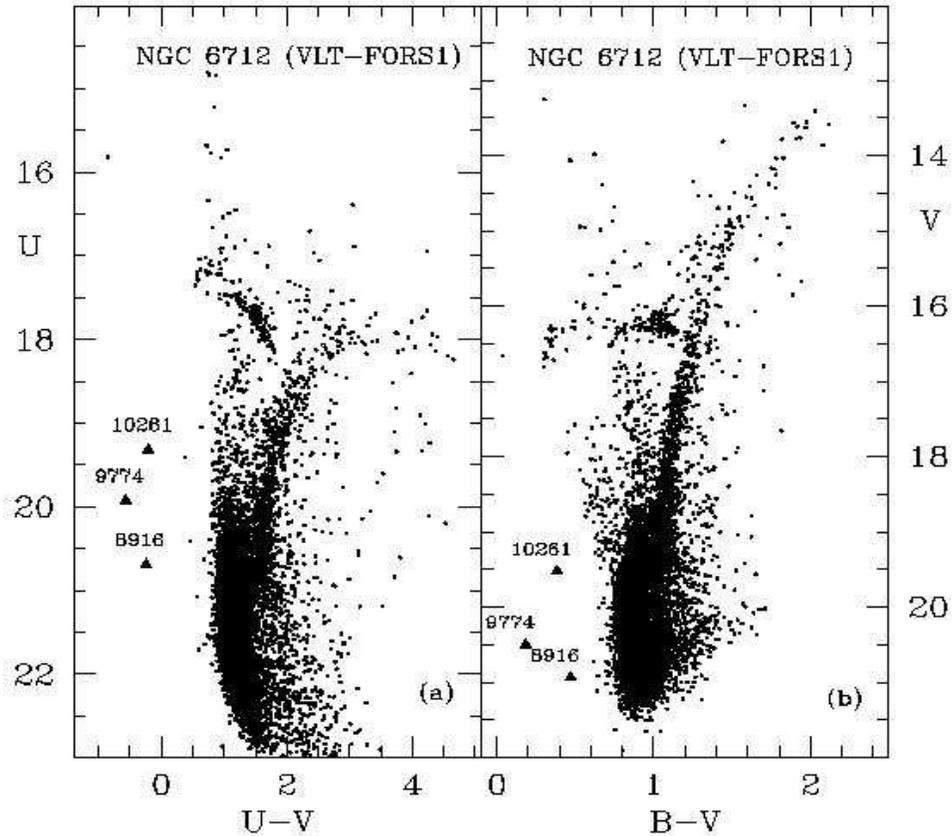}
\caption[fig1.ps]{
{\it Panel (a):} $(U,U-V)$ color--magnitude diagramme (CMD) for
NGC\,6712. All stars detected in the FORS1-HR field of view have been
plotted.
{\it Panel (b):} $(V,B-V)$ CMD for NGC\,6712 from  FORS1-HR images.
The three UV-excess stars have been plotted as large filled triangles
and marked with their identification number in  our catalogue.
\label{fig:map}}
\end{figure*}

\section{Observations and results }
  
The data have been obtained on 1999 June 16 at the {\it ANTU}  Unit
Telescope 1 (UT1) of the Very Large Telescope (VLT) at ESO on Cerro
Paranal (Chile) using FORS1.  A large set of frames were secured in the
central region of NGC\,6712, using  different levels of resolutions.
Here we discuss only the results obtained from short multiband
exposures using the High Resolution (HR) mode of FORS1. In this
configuration the plate-scale is $0\farcs1/pixel$ and the FORS1
$2048\times2048$ pixel$^2$ array has a global field of view of
$3\farcm4 \times 3\farcm4$.  The data consist of five 10\,s $B$-, $V$-, $R$-band
exposures, five 120\,s $U$-band exposures and one 700\,s $H\alpha$ exposure,
roughly centered on the cluster center. All the observations were
performed in service mode under good seeing conditions
(FWHM$=0\farcs4-0\farcs5$).

A more detailed description of the observations and reductions of the
data-set discussed in this Letter will be given elsewhere (Paltrinieri
et al. 2000). In short, all the reductions have been
carried out using ROMAFOT (Buonanno et al. 1983), a package
specifically developed to perform accurate photometry in crowded
fields. Independent searches have been performed in the best blue
($U,B$) and in the best red ($V,R$) images, in order to properly
optimize the search for blue and red objects.  The {\it blue} and the
{\it red} masks with the star positions have been then adapted to each
individual stellar image and the PSF-fitting procedure performed.  The
{\it blue} and the {\it red} data-set were then matched  together and a
final catalogue with the average instrumental magnitude in each filter
and coordinates has been compiled for all the stars identified in the
FORS1 field of view.

Photometric calibration of the instrumental magnitude was then
performed using ten photometric standard stars in selected areas
PG1528, PG2213, PG2331 (Landolt 1992).  $H\alpha$ magnitudes were
treated and calibrated like $R$-band images, except that an offset of
$\sim 4 $ mag was applied in order to account for the relative filter
efficiency.

\begin{figure*}[htb]
\vskip4truein
\includegraphics{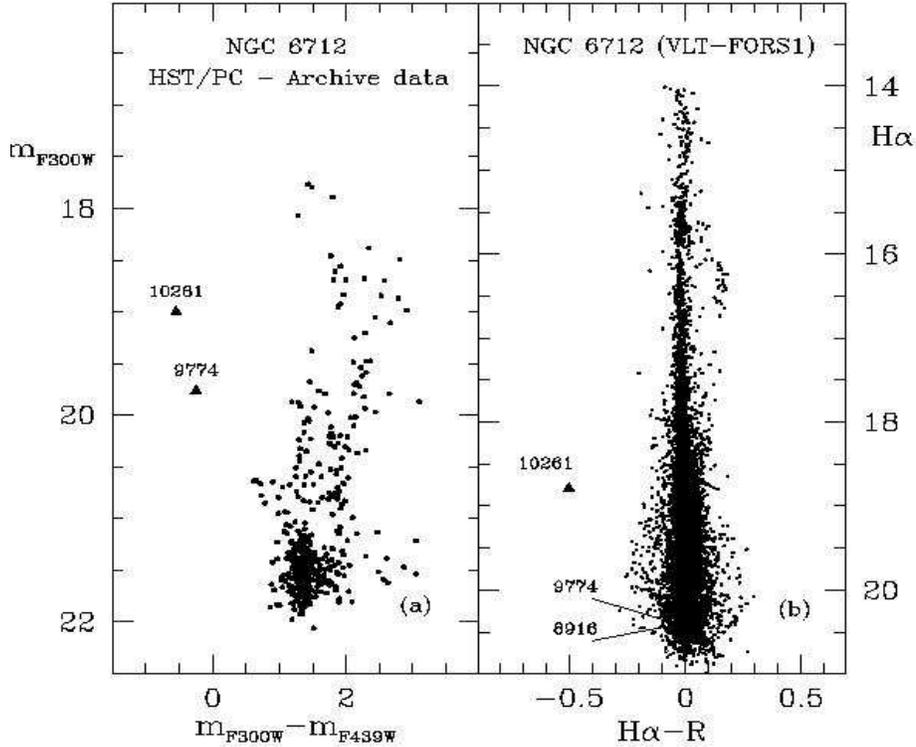}
\caption[fig2.ps]{
 {\it Panel (a):} ($m_{F300W}, m_{F300W}-m_{F439W}$)- CMD from HST-WFPC2
archive data. Only stars lying in the Planetary Camera are plotted.
{\it Panel (b):}  $(H\alpha, H\alpha-R)$ CMD for NGC\,6712 from FORS1-HR
images. Only star  $\#10261$ ({\it large filled triangle}) shows a
significant $H\alpha$ excess.  In both panels the UV-excess stars have
been plotted as large filled triangles and marked with their
identification number in our catalogue.
\label{fig:map}}
\end{figure*}

Figure 1 shows the $U,U-V$ CMD for more than 10,000 stars measured in
the FORS1-HR field of view.  As can be seen, apart from one bright blue
object at $U\sim 15.8, U-V\sim 0.8$ (a cooling post-AGB star?),
which will be discussed elsewhere (Paltrinieri et al. 2000, in
preparation), only three relatively faint ($U>18$) UV-excess stars
($U-V<0$) have been found to significantly lie outside the main loci
defined by the cluster stars. These stars ($\#10261$, $\#9774$ and
$\#8916$) are marked as large filled triangles in Figure 1.  The
peculiar blue colour of these stars is confirmed
by the $V, B-V$ CMD (Figure\,1\,b): where the three stars also
clearly lie outside the sequences defined by cluster stars.
 
In order to further  confirm the anomalous blue colours of the three
stars plotted in Figure\,1, we retrieved public WFPC2-HST data from the
archive. Four 300\,sec exposures through the F330W filter and one
160\,sec exposure in F439W, have been analysed. The standard procedure
described in Ferraro et al. (1997a) was used to reduce these
HST-images. The $(m_{F330W}-m_{F439W}, m_{F330W})$ CMD is plotted in
Figure 2(a), where only stars in the {\it Planetary Camera} (PC) have
been plotted.  As can be seen, both star $\#10261$ and $\#9774$ are
confirmed to be UV-excess object, while star $\#8916$ is well outside
the field of view of the PC.
 
A further test of the nature of these UV-excess stars can be made by
checking whether they exhibit $H\alpha$-emission.  Figure\,2\,b shows
the $(H\alpha-R, H\alpha)$ CMD.  Only unsaturated stars in the long
$H\alpha$ exposure have been plotted.  The CMD reveals that only star
$\#10261$ (among 10,000 stars measured) shows a significant excess of
$H\alpha$ emission. The other two UV-excess stars have {\it normal}
colour ($(H\alpha-R)>-0.1$), and thus they are fully compatible, within
the errors, with the $(H\alpha-R)$ colour of {\it normal} cluster MS
stars.

\begin{figure*}[htb]
\vskip4.7truein
\includegraphics{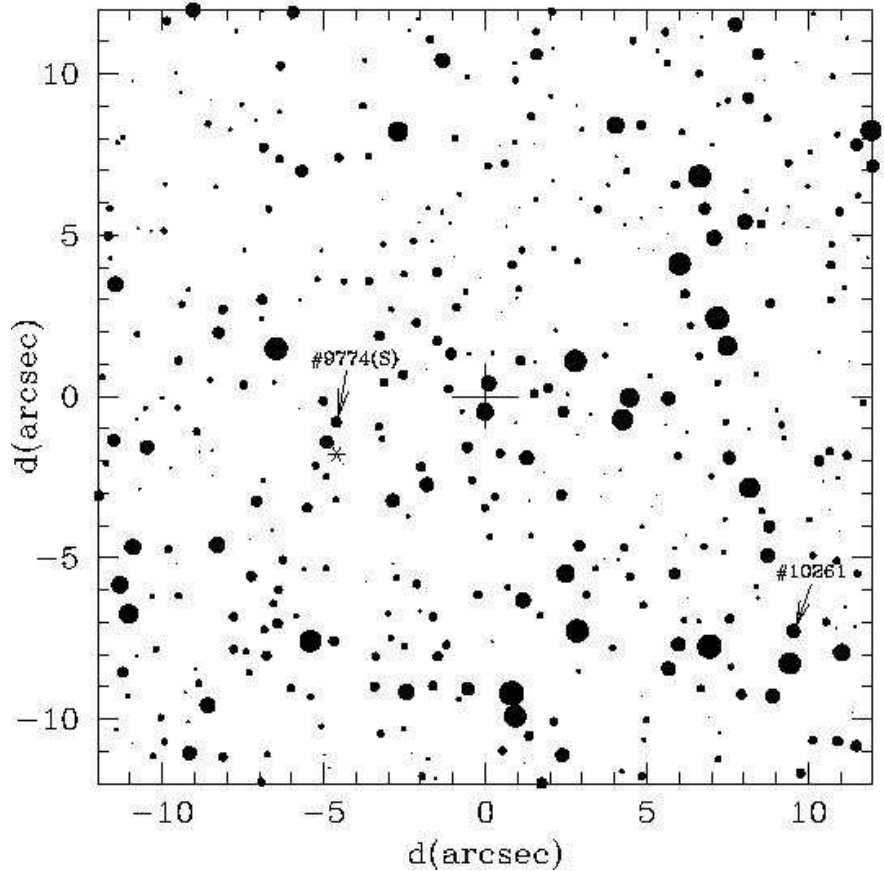}
\caption[fig3.ps]{
U map of the inner ($\sim 24'' \times 24''$)
region of the cluster.  The X and
Y scales are in arcsec with respect to the  center of gravity
($C_{grav}$ indicated by a cross in the figure).  The location of the
two UV-excess stars in the core is also marked. The nominal position of
the of the X-ray source is indicated by a  an asterisk.  North is up
and East on the left.}
\end{figure*}

\section{Properties of the UV-excess stars}

The absolute magnitudes and colours of the three UV-excess stars
have been computed adopting a distance modulus of $(m-M)_0=14.56$
and $E(B-V)=0.33$ (see the discussion in Paltrinieri et al. 2000).
Both distance modulus and reddening estimates have been
obtained applying the method described in Ferraro et al. (1999a).
  
In order to properly locate the three UV-excess stars with respect to
the cluster centre, we have determined the gravity centre ($C_{grav}$)
of the cluster following the procedure already adopted in other papers
(Ferraro et al. 1999b):  we averaged the X and Y pixel coordinates of
all stars with $V<20$ and lying within 1000 pixel ($\sim 100''$) of the
cluster centre, at first extimated by eye. $C_{grav}$ turns out to be
located at $\alpha_{\rm J2000} = 18^{\rm h}\, 53^{\rm m}\, 04\fs 6,
\delta_{J2000} = -08\arcdeg\, 42\arcmin\, 18\farcs 50$ which is $\sim
6\arcsec$ NE of the luminosity centre reported in the Djorgovski (1993)
compilation.
  
Grindlay et al. (1984) provide an accurate position (within a few
arcsec) of the LMXB derived from Einstein HRI at $\alpha_{\rm J2000} =
18^{\rm h}\, 53^{\rm m}\, 04\fs 91, \delta_{J2000} = -08\arcdeg\,
42\arcmin\, 20\farcs 15$.  More recently, Lehto et al. (1990)  detected
a radio emission from the X-ray source in NGC\,6712 and their accurate
Very Large Array position suggests that the global uncertainty on the
position of the X-ray source can be hardly larger than a few arcsec.

Absolute positions,  relative distances with respect to the $C_{grav}$
and to the  X-ray source, and the absolute magnitudes of the three
UV-excess  stars marked in Figure\,1 and 2 are listed in Table\,1.
 With the exception of star $\#8916$, which  is located $\sim 84''$ away
from the cluster center, well outside the cluster core ($r_c=56''$
Harris 1996), the other two objects (namely $\#9774$ and $\#10261$) are
very close to each  other and within a few arcsec of the cluster centre
and of the nominal position of the X-ray source (see Figure\,3).


Star $\#9774$ is star $S$ identified by Anderson et al. (1993) as the
optical counterpart to the LMXB.  
Its positions (see Table 1) turns to be
only $\sim 1''$ away from  the X-ray
source (in agreement with Anderson et al. (1993)). 
 Our observations confirm that it is the bluest object within
$\sim 15''$ of the X-ray source's nominal position, and for this reason
it remains the best candidate to be the optical counterpar to the
LMXB.  Following this assumption, we can use the B magnitude obtained
from our photometry to derive the parameter $\zeta = B_0+2.5 Lg F_X
(\mu J)$.  This parameter has been used by Van Paradjis \& McClintock
(1995) to characterize the ratio of X-ray to optical flux for LMXB in
the field, and recently used by Deutsch et al. (1998) to derive the
mean characteristics of LMXBs in GGCs.  Assuming for the reddening the
figure quoted above, we obtain $B_0=19.33$, and with the X-ray flux
listed in van Paradjis (1995), we obtain  $\zeta = 21.44$, a value
fully compatible with the mean value obtained by van Paradjis \&
McClintock (1995) for LMXBs in the field $<\zeta> = 21.8\pm 1.0$.
 
\begin{table*}[htb]
\caption{Position of the UV-excess stars in   NGC\,6712}
\begin{center}
\begin{tabular}{lllllllccc}
\tableline
\tableline
 Name & $M_U$ & $M_B$& $M_V$& $M_R$& $\alpha_{2000}$   & $\delta_{2000}$  & 
$d(center)$ & 
 $d(X-ray source)$ & \\
\tableline
$ X1850-086$ & -- & -- & -- & -- &$ 18^{\rm h}\, 53^{\rm m}\, 04\fs 91$ 
& $-08\arcdeg\, 42\arcmin\, 20\farcs 15$ &  $6\farcs$ & $ --$\\
$\#9774 (S)$ & 3.7 & 4.7 & 4.9 & 5.1 & $18^{\rm h}\, 53^{\rm m}\, 04\fs 91 
$ & $-08\arcdeg\, 42\arcmin\, 19\farcs 35$ &  $ 4\farcs 7 $ & $ 1\farcs  $\\
$\#10261$ & 3.1 & 4.0 & 3.9 & 4.0  &$18^{\rm h}\, 53^{\rm m}\, 03\fs 96$ & 
$-08\arcdeg\, 42\arcmin\, 
25\farcs 84$ &  $12\farcs$ & $ 15\farcs  $\\
$\#8916$ & 4.5  & 5.5 & 5.3 & 5.2 & $18^{\rm h}\, 53^{\rm m}\, 03\fs 44$ & 
$-08\arcdeg\, 40\arcmin\, 
57\farcs 70$ &  $83\farcs$ & $ 85\farcs  $\\
\tableline
\end{tabular}
\end{center}
\end{table*}

\section{Discussion}
  
Star $\#10261$ is the brightest object among the three faint UV-sources
marked as large triangles in Figure 1 and Figure 2. Moreover it is the
only object  showing a clearcut $H\alpha$ excess.  Star $\#10261$ is,
however, located $\sim 17''$ away from the X-ray source:  this distance
seems to be too large to  suggest a physical connection between this
object and the X-ray source.
                                                
$H\alpha$ emission and UV-excess detected for star $\#10261$, however,
suggest that this object might be a binary system in which a compact
object (neutron star or white dwarf) is accreting material  from a
secondary star via an accretion disc (the accretion disc being the
primary source of the UV emission and of the $H\alpha$ emission lines;
see Robinson 1976).
 
Thus star $\#10261$  is an additional excellent candidate LMXB in
quiescence (since quiescent LMXBs  show strong $H\alpha$ emission -
Grindlay 1994) or a CV. Although its colours are fully compatible with
the typical values obtained for CVs in the field:  ($V-R=0.5$,
Echevarria \& Jones 1984, $U-B=-1.0$), its absolute magnitude (see
Table\,1) is somewhat bright for a CV in quiescence. It is still
compatible with the magnitude of the accretion disc of a dwarf nova at
maximum of outburst ($M_V=3-6$, Warner 1987).  On the other hand, the
fact that bright CVs in outburst are expected  not to show strong
$H\alpha$ emission lines (Grindlay, 1994) works against this scenario.
 
X-ray to optical flux ratio is an important clue in disentangling CVs
and quiescent LMXBs (Grindlay 1984).  Thus, to shed some light on the
true nature of this object, we searched for possible X-ray emission in
this area. The images retrieved from the ROSAT archive, however, do not
show any secondary X-ray emission at the location of star $\#10261$.
Yet, this object is so close to star $S$ (see Figure\,3), that high
resolution, high sensitivity X-ray observations are needed to reveal
possible X-ray emission from this object, since the expected X-ray flux
from a CV or a quiescent LMXB is significantly lower (up to 4 orders of
magnitude) than that of an X-ray buster:  if a dim X-ray source were
connected to $\#10261$, it would be totally overwhelmed by the X-ray
emission from the burster.
 
LMXBs have been found to be  very overabundant (by a factor of $\sim
100$) in GGCs with respect to the field, where they are expected to
form by the evolution of primordial binaries. This finding is
generally interpreted as  evidence that the high stellar density has
led to many capture binaries.  On the other hand NGC\,6712,
 has a
probability to form LMXB via two body encounters which is $\sim
10^5-10^6$ times smaller that high-density clusters (Bellazzini et al.
1995). Thus the discovery of 
another by-product
of the evolution of a binary system in this low density cluster
increases the mystery about the formation of these objects.

Another possible by-product of binary systems evolution has been
recently serendipitously discovered (Deutsch et al. 1999)  a few
arcsec  away from the brightest X-ray source ever observed in a GGC, in
the core of NGC6624. We note here, however, that the discovery of
multiple binary systems in the vicinity of the very small core ($r_c
\sim 4''$) of the high-density ($Log \rho_0 \sim 5.6$), post-core
collapsed NGC6624 is   much less surprising   than what we have found
in the central regions of NGC\,6712, since the internal  dynamical
evolution of the cluster surely    plays a fundamental rule in the
formation of these objects. In fact, there is now a growing consensus
that dynamical evolution of the cluster can produce, via collisions,
several classes of peculiar objects (see, for example the exceptionally
large population of collisional blue stragglers recently discovered in
the core of the high-density GGC M80 - Ferraro et al. 1999a).

On the other hand, NGC\,6712  is not the only low-density   GGCs in
which candidate interacting binary systems have been found:  for
example, two  faint UV-excess stars, probably connected to low luminous
X-ray emission have been recently discovered (Ferraro et al. 1997b)
through HST observation in the core of M\,13, a  GGC with central
density $Log \rho_0 \sim 3.4$, comparable to that of NGC\,6712.  These
observations suggest that interacting binaries (LMXBs and CVs) are also
present in low-density GGCs, where they might originate from
the normal evolution of primordial binaries (see Davies 1997).
 
Furthermore, evidence is now mounting that the dynamical history of
NGC\,6712 is much more complex than that of other normal GGCs.  The
orbit computed by  Dauphole et al. (1996) for this cluster  suggests
that it is experiencing a strong interaction  with the disc and the
bulge of the Galaxy. This interaction might have deeply modified the
structure  of this cluster, over the last 10\,Gyr. The inverted mass
function found by De~Marchi et al.  (1999) and recently confirmed by
Andreuzzi et al. (2000), is a clear signature of strong tidal
stripping.  Moreover, Takahashi \& Portegies Zwart (2000), from
extensive N-body simulations, have recently found that such a
signature  is very rare and only appears in clusters which have lost
$\geq 99\%$ of their initial mass.  These facts suggest that NGC\,6712
was much more massive  and, probably, much more concentrated in the
past than it is today (Grindlay 1985 suggested it is a post collapsed
core in re-expansion phase), since  its dynamical evolution is driving
it towards dissolution.  If this scenario finds further support in the
observations, then we could conclude that what we are observing now is
only the {\it fossil remnant} core of one of the most massive cluster
in the Galaxy, may be even more massive than $\omega$ Cen.  Stellar
interactions among its stars, could have produced, at some earlier
epoch, a variety of exotic objects such as interacting binaries. Due to
mass segregation, some of these high-mass by-products are still
confined in the most inner region of the dissolving core (Takahashi \&
Portegies Zwart 2000) and might reveal their existence in the form of
the peculiar objects that we have discussed in this Letter.

\acknowledgments

We warmly thanks Luigi Stella and Gianluca Islael for useful comments
on the first draft of the paper.  We are also grateful to Wolfgang
Voges for making the ROSAT archive images of NGC\,6712 available to us
and to Enrico Vesperini and Oleg Gnedin for  many useful discussions on
the dynamical evolution of Galactic clusters.  The financial support of
the Agenzia Spaziale Italiana (ASI) and  of the {\it Ministero della
Universit\`a e della Ricerca Scientifica e Tecnologica} (MURST) to the
project {\it Stellar Dynamics and Stellar Evolution in Globular
Clusters} is kindly acknowledged.  F. R. F. gratefully acknowledges the
hospitality of the {\it Visitor Program} during his stay at ESO, when
most of this work has been carried out.

\end{document}